\documentclass[aps,prx,reprint, longbibliography,noeprint]{revtex4-2}

\usepackage{lipsum}

\usepackage{siunitx}
\DeclareSIUnit\gauss{G}
\usepackage{amsmath,bm}
\usepackage{dsfont}
\usepackage{amsbsy}
\usepackage{amssymb}
\usepackage{mathptmx, textcomp}
\usepackage{color}
\usepackage{braket}
\usepackage{graphicx}
\usepackage{textcomp}
\usepackage{notes2bib}
\usepackage{textcomp}
\usepackage{mwe}
\usepackage{siunitx}
\usepackage{physics}
\usepackage{filecontents}
\usepackage{soul}
\definecolor{bl}{rgb}{0, .1, .6}
\usepackage[colorlinks=true, citecolor = bl, linkcolor = bl, urlcolor=bl, pdfborder={0 0 0}]{hyperref}

\usepackage{changes}
\definechangesauthor[name=Igor, color=teal]{Igor}

\usepackage[letterpaper,top=2cm,bottom=2cm,left=3cm,right=3cm,marginparwidth=1.75cm]{geometry}

\usepackage{amsmath}
\usepackage{graphicx}

\definecolor{giu}{rgb}{1,0.7,0.7}

\begin{document}
\title{Single-atom resolved collective spectroscopy of a one-dimensional atomic array}
\author{Britton Hofer}
\altaffiliation{These authors contributed equally}
\author{Damien Bloch}
\altaffiliation{These authors contributed equally}
\author{Giulio Biagioni}
\author{Nathan Bonvalet}
\author{Antoine Browaeys}
\author{Igor Ferrier-Barbut}
\email{igor.ferrier-barbut@institutoptique.fr}
\affiliation{Universit\'e Paris-Saclay, Institut d'Optique Graduate School, CNRS, 
Laboratoire Charles Fabry, 91127, Palaiseau, France}

\begin{abstract}
Ordered atomic arrays feature an enhanced collective optical response compared to random atomic ensembles due to constructive interference in resonant dipole-dipole interactions. One consequence of this is the existence of a large shift of the transition with respect to the bare atomic frequency. In the linear optics regime (low light intensity), one observes a spectroscopic shift of the Lorentzian atomic line often called the collective Lamb shift. For stronger driving, many excitations are present in the system rendering the calculation of this shift theoretically challenging, but its understanding is important for instance when performing Ramsey spectroscopy in optical clocks. 
Here we report on the study of the collective optical response of a one-dimensional array of 30 dysprosium atoms. We drive the atoms on the narrow intercombination transition isolating a 2-level system, and measure the atomic state with single-shot state readout using a broad transition. In the linear optics regime, we measure the shift of the resonance in steady state due to dipole interactions, and measure how this shift depends on the interatomic distance. We further resolve at the single atom level how the excitation is distributed over the array. Then, on the same transition we perform Ramsey spectroscopy \emph{i.\,e.}~away from the linear regime. We observe a time-dependent shift, that allows us to draw the connection between the collective Lamb shift observed in the linear optics regime and in the large-excitation case.
\end{abstract}

\maketitle

\section{Introduction}
Collective light-matter interactions in atomic ensembles are an example of a dissipative quantum many-body problem that has been studied for decades, both theoretically and experimentally \cite{Dicke1954,Gross1982,Allen1987}. Recent progress in atomic physics has led to new investigations in different directions. Particular experimental attention has been devoted to the understanding of the shift of an atomic transition due to resonant dipole-dipole interactions, the so-called collective Lamb shift \cite{Friedberg1973}. On the one hand, many works have probed the shift of a Lorentzian line \cite{Meir2014,Bromley2016,Jennewein2016,Corman2016,Peyrot2018,Glicenstein2020,Skljarow2022,Vatre2024,Scully2009,Rohlsberger2010,Roof2016,Ido2005,Javanainen2014,Zhu2016}, in the regime of \emph{low} light intensity where the atomic dipoles behave linearly and can be considered as classical. On the other hand, it is important to understand this shift in the context of optical clocks \cite{Chang2004,Cidrim2021,Hutson2024}, where one typically performs Ramsey spectroscopy, \textit{i.\,e.}~outside the linear regime, where the non-linear response of single atoms cannot be ignored. In addition, in such systems, the atoms are usually ordered. The influence of geometrical order on the collective response of an atomic ensemble to light is now the topic of an intense research activity 
\cite{Chang2004,Kramer2016,Facchinetti2016,Facchinetti2016b,Bettles2016,Bettles2016,Shahmoon2017,Asenjo2017,Glicenstein2020}, and the enhanced collective response of two-dimensional arrays in the linear optics regime has been investigated experimentally \cite{Rui2020,Srakaew2023}. \par

In this context, we present here an experiment on a tweezer-based one-dimensional array of two-level atoms, where we develop a single-atom resolved single-shot state readout technique. As a first demonstration of the possibilities it offers, we use it to probe the collective shift, both performing spectroscopy near the linear optics regime \emph{and} Ramsey spectroscopy in the non-linear regime: where atoms act as quantum emitters and not linear dipoles. 
In the low saturation regime, we measure the collective shift in steady state for different atomic spacings. We then reveal how the excitations spread along the array, measuring for the first time the microscopic effect of resonant dipole-dipole interactions at the single atom level. As the strength of the drive increases, the steady state shift disappears so we probe the interactions in the non-linear regime by performing Ramsey spectroscopy with a duration on the order of the excited state lifetime. This regime allows us to experimentally show how the clock shift in Ramsey spectroscopy evolves in time and how it is related to the line shift in the linear regime.


\section{Experimental setup}
To probe the collective Lamb shift in an ordered one-dimensional system, we prepare arrays of single dysprosium (Dy) atoms using the experimental platform presented in \cite{Bloch2023,Bloch2024}. The atoms are held in optical tweezers with a wavelength of $\SI{532}{nm}$. We produce ordered arrays of 30 atoms, with a controllable inter-tweezers spacing $d$ from $\SI{1.25}{\micro\meter}$ to $\approx\SI{4.5}{\micro\meter}$. These arrays are obtained by rearranging a randomly loaded chain of 75 tweezers \cite{Endres2016}. \par
\begin{figure*}[t]
    \includegraphics[width=445pt]{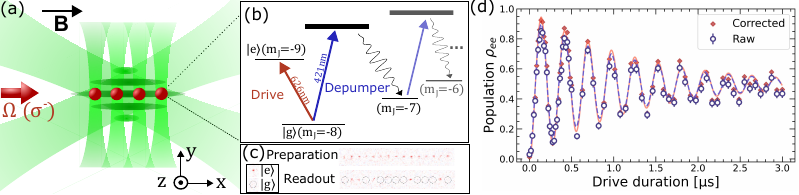}
    \caption{\textbf{(a)} Schematic of the experiment: We perform spectroscopy on an array of Dy atoms held in optical tweezers of wavelength \SI{532}{nm} and further confined along the weak trapping axis of the tweezers with a shallow-angle optical lattice, also at \SI{532}{nm}. The red arrow shows the propagation direction of the driving laser with which we excite the atoms. 
    \textbf{(b)} Dysprosium's optical transitions used in this work. Spectroscopy is performed on the \SI{626}{nm} intercombination line between $\ket{g}=\ket{m_J=-8}$ and $\ket{e}=\ket{m_J=-9}$ (linewidth $\Gamma=(2\pi)\,\SI{135}{kHz}$). The broad \SI{421}{nm} line ($\Gamma_{421}=(2\pi)\,\SI{32.5}{MHz}$) is used to perform single-shot state readout by rapidly depumping atoms out of $\ket g$. \textbf{(c)} Readout of the atomic state in a single shot. The atom is reimaged if it was shelved in $\ket{e}$ and atoms in $\ket{g}$ were depumped to higher $m_J$ states and are thus not re-imaged. \textbf{(d)} Measurement of the excited state fraction during Rabi oscillations for independent atoms, using the shelving technique illustrated in (b). The blue circles are raw data, while the red diamonds include corrections for detection errors (see appendix \ref{appendix:readout}).  Error bars represent standard error on the mean. The lines represent solutions of the optical Bloch equations.
 }
    \label{fig:figure1}
\end{figure*}
In the tweezers, the radial trap frequency is $\omega_r/(2\pi) = \SI{50}{kHz}$ while the axial one is $\omega_z/(2\pi) = \SI{7}{kHz}$. Given the atomic temperature in the tweezers $T=\SI{5.5}{\micro K}$, this yields a radial positional disorder of size $\sigma_r\simeq\SI{50}{nm}$ and an axial one of $\sigma_z\simeq\SI{400}{nm}$. This axial disorder is non-negligible with respect to the interatomic distance and hinders collective effects. In order to reduce disorder, we add an extra confinement along the tweezers' axis using an optical lattice produced by two $\SI{532}{nm}$ beams interfering at a half angle of about $\SI{5}{\degree}$, yielding a lattice spacing of $\SI{3}{\micro m}$ (see Fig.~\ref{fig:figure1}). All atoms are loaded in a single lattice plane \cite{Young2022}, perpendicular to the tweezers' axis. We give details in appendix \ref{appendix:axial_trapping}. Adiabatically turning this lattice on increases the axial trapping frequency up to $\omega_z/(2\pi)=\SI{35}{kHz}$ at the cost of an increased temperature of $T=\SI{8.5}{\micro K}$, resulting in final sizes $\sigma_r\simeq\SI{70}{nm}$ and $\sigma_z\simeq\SI{90}{nm}$ in these near-spherical traps.\par

Having prepared this 1D ordered system, we perform spectroscopy on the intercombination transition of $^{162}$Dy with wavelength $\lambda=\SI{626}{nm}=2\pi/k$ and linewidth $\Gamma=(2\pi)\,\SI{135}{kHz}$ which connects a $J=8$ to a $J=9$ Zeeman manifold [excited state $4f^{10}(^5I_8)6s6p(^3P^\circ_1)\,(8,1)^\circ_9$]\cite{NIST_ASD}. We isolate a two-level transition between $\ket g=\ket{J=8,m_J=-8}$ and $\ket e=\ket{J=9,m_J=-9}$ by applying a magnetic field of $\SI{7}{G}$ and tuning the driving laser to the $\sigma^-$ transition frequency (the $\pi$ transition is detuned by about \SI{13}{MHz}). The magnetic field is along y during the imaging so as to have magic trapping due to an elliptical polarization \cite{Bloch2023,Bloch2024}. We then turn the magnetic field to be along x during the light scattering experiment in a few milliseconds. Because the polarization of the drive is $\sigma^-$, we choose the direction of the magnetic field which maximizes the radiation of the atoms along the chain. In these conditions the tweezers and the lattice (which is linearly polarized) are non magic. 
As a consequence all experiments reported here are performed in free space (the tweezers and the lattice are turned off) to avoid inhomogeneous light shifts. The residual atomic motion during the free flight is negligible ($<0.2\,\lambda$). In all the experiments we report, the light scattering sequence takes place on a time-scale shorter than $\SI{10}{\micro s}$ such that we recapture the atoms with a high probability \cite{Bloch2023}.\par

Nearly all studies of light scattering in free space rely on measurements of the radiated light. However, the theoretical description of this problem computes the many-body atomic density matrix, and the radiated field is calculated as a linear superposition of the atomic dipoles \cite{Kiffner2010}. As such it is interesting to directly access the atomic state \cite{Glicenstein2024}. Here we use a method based on shelving $\cite{Nagourney1986,Sauter1986}$  from a broad transition [excited state $4f^{10}(^5I_8)6s6p(^1P^o_1)\;(8,1)^o_9$, wavelength $\SI{421}{nm}$, $\Gamma_{421}=(2\pi)\,\SI{32.5}{MHz}$] to readout the internal state. We apply an optical depumping pulse ($\tau_p=\SI{100}{ns}$) on the broad transition with $\pi$ polarization. This pulse depumps $\ket g$ atoms to other Zeeman states with $m_J>-8$. Atoms in $\ket e$ do not significantly decay in this interval as the lifetime of atoms in $\ket{e}$ is $\tau=1/\Gamma=\SI{1.2}{\micro s}$. After the pulse, $\ket e$ atoms decay to $\ket g$. We re-image only the atoms left in $\ket g$, while those in the other Zeeman states are not imaged (more details in appendix \ref{appendix:readout}). With this, we project the atomic state and image only the atoms that were in $\ket e$ before the blue depumping pulse. We thus obtain state readout (\emph{i.e.} measurement of $\rho_{ee}$) of single atoms in a single shot with a fidelity limited by the excited state lifetime $F=0.92\simeq e^{-\tau_p/\tau}$. This method is in principle lossless if the atoms are efficiently repumped to $\ket g$ at the end of an experimental cycle. This would greatly improve the duty cycle, and we leave its implementation to future works. Fig.~\ref{fig:figure1}(c) shows an example of Rabi oscillations measured with this method.\par

\begin{figure}[b]
    \includegraphics[width=\columnwidth]{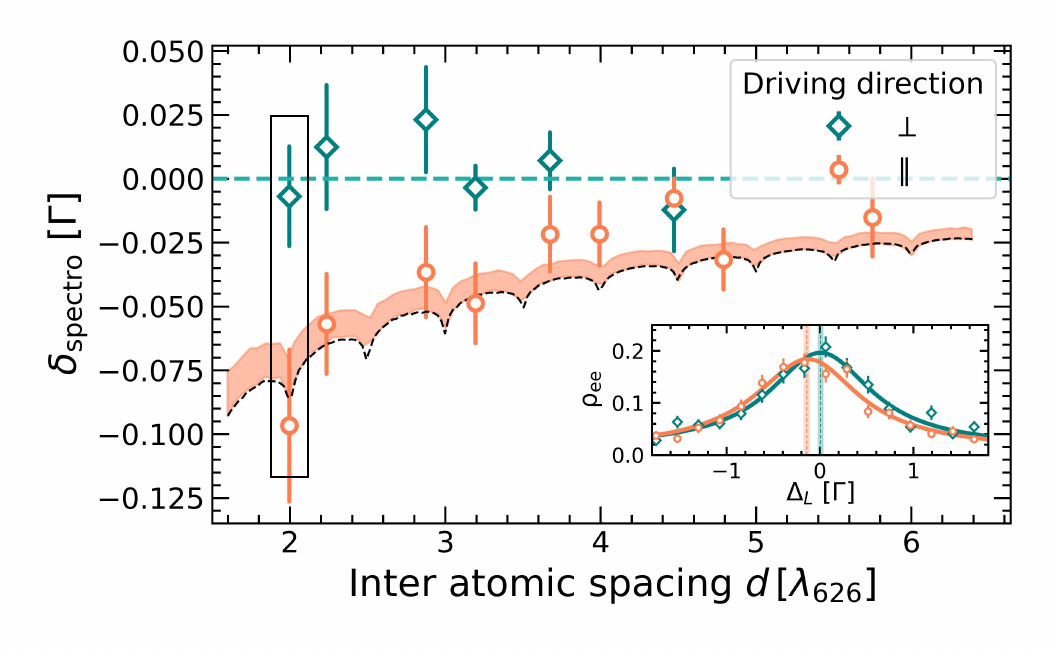}
    \caption{
    Collective frequency shift of a 1D chain of 30 atoms when the atomic spacing is varied. The Rabi frequency of the drive is $\Omega=0.80(8)\Gamma$. The orange circles correspond to a drive along the chain and the green diamonds to a perpendicular drive. Error bars represent the standard error on the mean. When the driving laser is sent parallel to the chain, a frequency shift is observed that increases as the atoms get closer. In solid orange is the result of mean-field simulations accounting for positional disorder due to temperature. The shaded area accounts for the 10\% uncertainty of the Rabi frequency of the drive. We include the analytical prediction of the shift [Eq.\eqref{eq:delta_spectro}], in black dashed line. 
    The inset shows the measurement of the atomic line for the two points highlighted by the black box.}
    \label{fig:freq_vs_spacing}
\end{figure}

\section{Shift in steady-state in the low saturation regime}
To benchmark the experiment, we first measure the collective frequency shift ($\delta_\mathrm{spectro}$) of the atomic line due to interactions for different interatomic spacings \cite{Meir2014,Glicenstein2020}, Fig.~\ref{fig:freq_vs_spacing}. 
For this, we excite the atoms with a near-resonant laser, detuned by an amount $\Delta_L$ from the bare atomic resonance, with low intensity $I=1.3\,I_{\rm sat}$ \textit{i.\,e.\,}Rabi frequency $\Omega=(2\pi)\,\SI{110}{kHz}=0.8\,\Gamma$.  
This laser propagates either along the chain, or perpendicular to it. In the case where the drive runs parallel to the chain $\vec k_\mathrm{las}=k\,\hat x$, the propagation phases of the fields radiated by each atomic dipole are the same as the propagation phase of the drive $e^{i \vec k_\mathrm{las}\cdot \vec r}=e^{ikx}$. 
This results in an interference of the resonant dipole-dipole interactions along the chain and a shift of the atomic line \cite{Sutherland2016}.
At vanishing Rabi frequency, the line shift in steady state is given by the average interaction energy in the array \cite{Chang2004}: $\delta_\mathrm{spectro}^0=\frac{1}{N\,\hbar}\sum_{n\neq m}{\mathrm Re}[V_\mathrm{dd}(\vec r_n-\vec r_m)e^{-i\vec k_\mathrm{las}.(\vec r_n-\vec r_m)}]$ and $V_\mathrm{dd}$ is the resonant dipole-dipole interaction $\propto1/d$ at large distances for a $\sigma^-$ transition:
\begin{align*}
    V_\mathrm{dd}(\vec r)=-\frac{3\Gamma\hbar}{8}\frac{e^{iv}}{v}\left[\zeta^2+1+(3\zeta^2-1)\left(\frac{i}{v}-\frac{1}{v^2}\right) \right]
\end{align*}
Where $v=kr$, $\zeta=\cos\alpha$ and $\alpha$ is the angle between the quantization axis of the dipoles and $\vec r$.
To ensure a measurable excitation fraction, our experiment is performed with a non negligible Rabi frequency ($\Omega=0.8\Gamma$), but relatively weak interactions. 
In appendix \ref{appendix:analytical_shift}, we derive the theoretical expression of the atomic line in the mean-field, weakly interacting case, Eq.\,\eqref{eq:line}. The line is a skewed Lorentzian, with a maximum shifted by:   
\begin{equation}
    \delta_\mathrm{spectro}=\frac{\delta_\mathrm{spectro}^0}{1+2\Omega^2/\Gamma^2}.
    \label{eq:delta_spectro}
\end{equation}
We show this analytical prediction as black dashed line in Fig.~\ref{fig:freq_vs_spacing}, averaging over positional disorder. 
To verify it, we perform time-dependent numerical simulations of the dynamics within a mean-field approximation \cite{Glicenstein2020}. They account for the time of the pulse and averaging over thermal positional disorder (see more details in appendix \ref{appendix:NLCD} on how we incorporate the atoms' finite temperature). 
The experimental data and the simulations are fit with the same. theoretical line profile [Eq.\,\eqref{eq:line}] leaving the shift as a free parameter, which is what is represented in Fig.\,\ref{fig:freq_vs_spacing}.
Both the analytical expression and full simulations show good agreement with the data. 
The theory curves display small oscillations with spacing $\lambda/2$ \cite{Meir2014}, which are not the focus of our work. In the case where the driving is perpendicular, the coherent buildup does not occur and only a weak shift is expected and measured. 
\par
\begin{figure}[tbp]
    \includegraphics[width=\columnwidth]{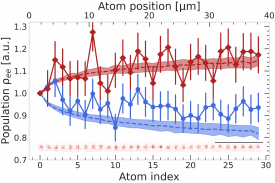}
    \caption{Position-resolved excited state fraction where the error bars represent the standard error on the mean. A driving beam detuned by $\Delta_L=\pm\Gamma/2$ is sent along the atomic chain from left to right. The atoms are $\SI{1.4}{\micro\meter}=2.2\,\lambda$ apart. The joined red (blue) diamonds (dots) are the experimental data for $\Delta_L=-\Gamma/2\,(+\Gamma/2)$, normalized to the fraction on the first atom. The dashed lines show mean-field simulation results averaged over thermal positional disorder. One observes a decay or growth of excitation along the chain depending on the detuning. The shaded area around the simulations accounts for a 10\% uncertainty of the Rabi frequency.}
    \label{fig:excitation_prob}
\end{figure}

As stated above, the resonance shift in a 1D system is due to a buildup of the interaction along the chain. At the single atom level, this should result in an excitation probability that varies along the chain \cite{Sutherland2016}. Thanks to single atom state readout, we can now directly measure how the excitation probability evolves along the chain. 
For a non-interacting system, it should be flat. 
However, we observe a different behavior visible in Fig.\,\ref{fig:excitation_prob}, where we plot the excitation probability of each individual atom in the chain. 
We present data for two different detunings, taken over several days, with different Rabi frequencies always around $\Omega=0.8 \Gamma$, such that the absolute excitation probability is meaningless. To compare it to theory, we thus normalize the excitation probability to that of the first atom in the chain. 
Since this first point is prone to statistical noise, this normalization results in a statistical offset of all the points. However, we still use it to keep the data analysis free of adjustable parameters (see more details in appendix \ref{appendix:data analysis}). When driven by a red-detuned probe ($\Delta_L<0$), each atomic dipole radiates a field which is in phase with the driving laser, leading to constructive interference with the drive. As the effective driving strength is increased along the laser's propagation, so is the excitation probability. We indeed observe an \emph{increase} of the excitation probability along the chain when $\Delta_L=-\Gamma/2$, in the direction of the drive's propagation. Conversely, for a blue-detuned probe ($\Delta_L>0$), the fields radiated by the atoms are out of phase with the drive, resulting in a decrease of the effective driving strength along the laser's propagation. In fact, when $\Delta_L=+\Gamma/2$ we observe a \emph{decrease} of the excitation probability along the chain. This effect was described in \cite{Sutherland_2017} and underlies observations reported in \cite{Glicenstein2020}, but for the first time we are able to resolve atomic excitation \emph{at the single atom level} to reveal this mechanism. One might also explain the enhancement of the atomic excitation on the red side of the resonance as the first atoms focusing the field, increasing its value on the atoms downstream, hence acting as an effective lens, seen here at the single atom level.\par

The results presented so far were obtained in the low light intensity regime where the atoms behave as classical, linear dipoles \cite{Ruostekoski1997}. Away from this limit, when many atoms are excited, one has to consider the full Hilbert space and non-trivial correlations should emerge at short interatomic distances \cite{Asenjo2017,Zhang2019,Henriet2019,Sheremet2023}.  Ramsey spectroscopy is an example where one operates far from the linear regime. Here, we explore in this simple system of 30 atoms how the shift measured above in the linear regime in steady state can be related to the shift observed in Ramsey spectroscopy. \par

\section{Non-linear regime}
\subsection{Steady state}

\begin{figure}[tp]
    \includegraphics[width=\columnwidth]{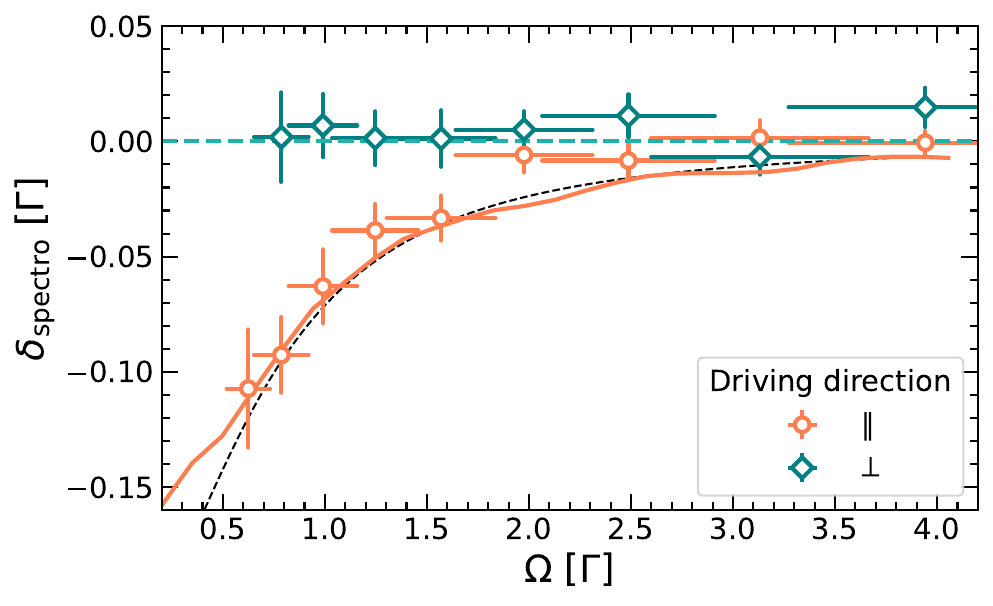}
    \caption{
    Frequency shift of the maximum of the excitation fraction measured in steady state, as a function of driving power for a fixed inter-atomic distance of $d=\SI{1.25}{\micro\meter}=2\,\lambda$. Error bars represent standard error on the mean. As the drive becomes too strong, the atomic dipole is progressively reduced causing the collective shift to vanish. The solid line is the result of mean-field simulations. The dashed black line is the trace of the analytical formula for the shift in the weakly interacting case.
    }
    \label{fig:freq_shift_vs_power}
\end{figure}

To depart from the linear optics regime, we first measure the line shift in steady state as above, simply increasing the frive Rabi frequency to $\Omega>\Gamma$. We fit the experimental points with a skewed Lorentzian line as discussed above and represent in Fig.~\ref{fig:freq_shift_vs_power} the fitted shift as a function of the Rabi frequency of the drive. The experimental data are still taken in steady state after driving for $\SI{7}{\micro s}\gg1/\Gamma$. It is acquired in a range of $\SI{500}{kHz}$ centered on the bare atomic frequency. We also perform mean-field simulations at various Rabi frequencies. When comparing with the experiment, we fit with the same function [Eq.\,\eqref{eq:line}] in a \SI{500}{kHz} window as is done with the experimental data. We find that if one does not take into account the distortion of the atomic line, the simulations do not reproduce the experimental data well. In the future, it would be interesting to explore the line distortion experimentally. The simulation results are represented as a solid line in Fig.~\ref{fig:freq_shift_vs_power}. The data is in very good agreement with the simulations and analytical expression Eq.\,\eqref{eq:delta_spectro} (black dashed line).
We observe, in agreement with our previous work \cite{Glicenstein2020}, that the shift is suppressed as soon as the Rabi frequency is on the order of $\Gamma$. As explained in \cite{Glicenstein2020}, the shift suppression is mainly due to the fact that the average atomic dipole ($\propto\rho_{eg}$) in steady-state vanishes like $\sim1/\Omega$.\par

\subsection{Ramsey spectroscopy}

On the other hand, Ramsey spectroscopy does not operate in steady state. The atomic dipoles are initialized to a non-zero value by a first pulse with duration $\ll1/\Gamma$ and area $\theta_0$. The dipoles then evolve without drive before the second pulse. The shift of the transition $\delta_\mathrm{Ramsey}$ can be calculated in a mean-field approximation, assuming that the time between the two pulses is very short ($T_\mathrm{Ramsey}\ll1/\Gamma$): in this case one obtains $\delta_{\rm Ramsey}=-\delta_{\rm spectro}^0\,\cos\theta_0$ \cite{Chang2004}. The shift of the Ramsey fringes depends on the amount of excitation $\rho_{ee}=(1-\cos\theta_0)/2$, as observed in \cite{Hutson2024}. This formula is valid only for $T_\mathrm{Ramsey}\ll1/\Gamma$. For longer times, the excitation decays due to spontaneous emission. As a consequence, one expects a \emph{time-dependent} shift for Ramsey interferometry in the regime $T_\mathrm{Ramsey}\sim1/\Gamma$. The time dependence of the shift was theoretically investigated for short times and with a pulse area $\theta_0 = \pi/2$ in ref.\,\cite{Chang2004}, longer times and other pulse areas were considered in \cite{Cidrim2021} (for a 3D ensemble). In the following, we derive an analytical formula which indeed predicts a time-dependent frequency shift and, to the best of our knowledge, experimentally observe this time-dependent shift for the first time.\par
\begin{figure}[b!]
\includegraphics[width=\columnwidth]{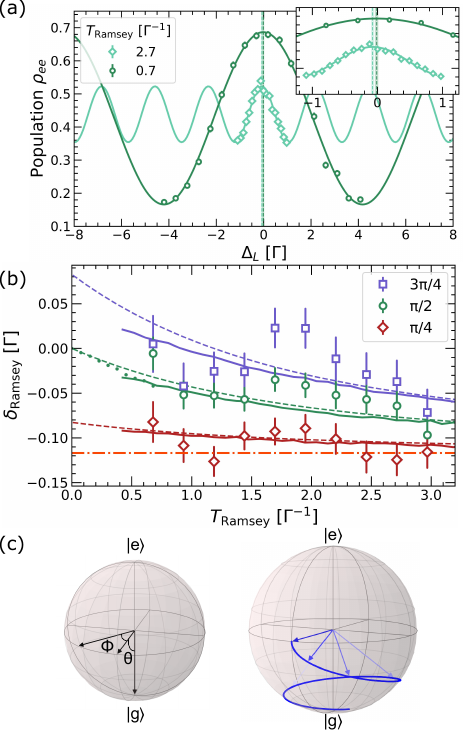}
    \caption{\textbf{(a)} Ramsey fringes with a fixed pulse area $\theta_0=\pi/2$ for two different $T_\mathrm{Ramsey}$ times corresponding to the free evolution time between the two pulses. \textbf{(b)} Time-dependent shift measured in Ramsey spectroscopy when changing the time $T_\mathrm{Ramsey}$ between the two pulses, for three different pulse areas $\theta_0$. For this experiment, the atoms are $\SI{1.4}{\micro\meter}=2.2\lambda$ apart and the drive's Rabi frequency is $\Omega=(2\pi)\,\SI{3}{\MHz}=22\,\Gamma$. Error bars represent standard error on the mean. Results of the mean field simulations are shown as solid lines. The dashed-dotted horizontal orange line is the predicted $\delta_\mathrm{spectro}^0$ shift calculated in the low-intensity regime with the mean field simulations. The data exhibits a time dependence in agreement with theory, connecting the large excitation regime of Ramsey spectroscopy to the low-excitation linear regime of fluorescence spectroscopy. The dotted green line shows the linear expansion for $\pi/2$ pulses from \cite{Chang2004} in the limit where $T_\mathrm{Ramsey}\Gamma\ll1$. We use an ansatz with an excitation-dependent instantaneous shift (see main text), plotted with dashed lines for each $\theta_0$. \textbf{(c)} Schematic representation of the origin of the time-varying shift. Bloch sphere representation of the state of an atom during the time evolution $T_\mathrm{Ramsey}$ between the two Ramsey pulses. Following a $\pi/2$ pulse the Bloch vector precesses around the $z$ axis at a varying rate: initially $\dot\phi=0$ while it reaches a non-zero value once the excitation has decayed, resulting in a shift which depends on the free evolution time.
    }
    \label{fig:ramsey}
\end{figure}

To investigate it, we perform a Ramsey spectroscopy experiment on the $\ket g\to \ket e$ transition with a chain of 30 atoms. The short pulses (area $\theta_0$) are created by a fiber electro-optic modulator with rise time $\sim \SI{1}{ns}$, and an additional AOM for good extinction. For this dataset, the atoms were not trapped in the lattice prior to the release in free-space \footnote{When using the lattice for measuring the shift of the Ramsey fringes, we observed systematic shifts which we ascribe to a slow turn-off of the inhomogeneous lattice potential. This inhomogeneous broadening plays a much stronger role in Ramsey interferometry in comparison to the slow steady-state experiments presented above. As each of the tweezers have the same intensity, they do not create inhomogenous broadening.}, which results in a lower shift of the transition.  We perform a simple sequence of two pulses separated by a time $T_\mathrm{Ramsey}$ as considered in \cite{Chang2004}. We pick three pulse areas: $\pi/4$, $\pi/2$ and $3\pi/4$. For each pulse area, we measure Ramsey fringes as a function of the laser detuning $\Delta_L$ (see Fig.~\ref{fig:ramsey}(a)) and we record the position of the central fringe. We repeat the experiments for different $T_\mathrm{Ramsey}$ times between $\SI{0.8}{\micro s}=0.7\,\Gamma^{-1}$ and $\SI{3.5}{\micro s}=2.7\,\Gamma^{-1}$, see Fig.~\ref{fig:ramsey}(b). We do not measure the shift of the Ramsey fringes for shorter Ramsey times because the fringe period is large ($>\SI{}{MHz}$) and these detunings become non-negligible with respect to the frequency difference with other transitions
\footnote{The probe beam has linear vertical polarization and propagates parallel to the magnetic field, it thus has mostly $\sigma^-$ and $\sigma^+$ polarization components.}, breaking the assumption of a two-level atom.\par 

Our observations are reported in Fig.~\ref{fig:ramsey}(b). Despite some spread around the main tendency \footnote{Some experimental shift might be due to systematic errors which are hard to model, such as a small frequency chirp in the fast-turn on of the AOM or a weak Doppler shift imparted on atoms during the interrogation. These systematics would distort the datasets identically, they are not accounted for in simulations.}, they clearly exhibit a dependence on the pulse area, as reported in \cite{Hutson2024}. The second striking feature is that the collective Lamb shift \emph{evolves in time}, with a slope that depends on the pulse area, well reproduced by mean-field solutions of the master equation (solid lines). For the smallest pulse area ($\pi/4$), we find a nearly time-independent shift with a value close to the expected spectroscopic shift in the linear case (dashed-dotted orange line). For larger pulse areas the shift decays in time in agreement with the qualitative explanation given above. We discuss the implications of these findings below. \par

The data of Fig.~\ref{fig:ramsey}(b) imply that the Bloch vector of an atom precesses at a frequency which varies during the time separating the two pulses. 
To illustrate this, let us take the case of $\pi/2$ pulses, with the laser having a zero detuning with respect to the atomic resonance. In this case the precession rate starts from the initial value of $\dot\phi(t=0)=-\delta_{\rm spectro}^0\,\cos\theta_0=0$, where $\phi$ is the phase of the Bloch vector (see Fig.~\ref{fig:ramsey}(c)), leading to a zero shift for short wait times ($T_\mathrm{Ramsey}\Gamma\ll1$).
As explained in ref.\,\cite{Chang2004}, this zero initial shift at $\theta_0=\pi/2$ can be formally understood writing the contribution of other atoms to the time evolution of an atomic dipole:  $ (\dot\rho_{eg})_\mathrm{int} =i\, \Omega_{\rm atoms}\left(1-2\rho_{ee}\right)/2$ where $\Omega_{\rm atoms}$ is the Rabi frequency of the field radiated by the other atoms (see Eq.\,\eqref{eq:OBE}). For a $\pi/2$ pulse, $\rho_{ee}=1/2$ and the dipole is not impacted by the others. Intuitively, this is due to the fact that during the free evolution time, the other atomic dipoles oscillate at the bare atomic frequency, and thus create an effective drive \emph{in the equatorial plane} of the Bloch sphere. This field thus does not drive a precession in the equatorial plane ($\dot\phi=0$) and does not lead to a shift of the transition for $\Gamma T_\mathrm{Ramsey}\ll1$. \par

For longer times ($T_\mathrm{Ramsey}\Gamma\gtrsim1$), the situation is different. As $\rho_{ee}$ decays, the dipole evolution starts to depend on the other atoms. The authors of \cite{Chang2004} derived a short-time linear expansion represented as a dotted line in Fig.~\ref{fig:ramsey}(b). Here, we extend the prediction of the shift to arbitrary time for weak interactions using the mean-field approximation, see appendix \ref{appendix:ramsey_shift}.
We find that the shift of the Ramsey fringes for a time $T_\mathrm{Ramsey}$ is given by 
\begin{equation}
\delta_\mathrm{Ramsey}=\delta_\mathrm{spectro}^0\left[1-\left(1-e^{-\Gamma T_\mathrm{Ramsey}} \right) \frac{1-\cos\theta_0}{\Gamma T_\mathrm{Ramsey}}\right].
\label{eq:Ramsey_shift}
\end{equation}
This shift can be understood by noting that the instantaneous precession rate is given by $\dot\phi(t=0)=-\delta_{\rm spectro}^0\,\cos\theta(t)$ where $\theta(t)$ follows the population decay due to spontaneous emission. Integrating this instantaneous precession rate to obtain the shift after a given time $T_\mathrm{Ramsey}$: $\delta_\mathrm{Ramsey}=\frac{1}{T_\mathrm{Ramsey}}\int_0^{T_\mathrm{Ramsey}}\dot\phi dt$ yields  Eq.\,\eqref{eq:Ramsey_shift}.\\

The result is represented as dashed lines in Fig.~\ref{fig:ramsey}(b), and reproduces the experimental data and mean-field simulations well. At long times ($\Gamma T_\mathrm{Ramsey}>1$ for which  $\theta\to0$), the precession rate approaches the value $\dot\phi=\delta_{\rm spectro}^0$. The evolution of $\delta_\mathrm{Ramsey}$ is schematically represented on Fig.~\ref{fig:ramsey}(c). Hence, the measured shift of Ramsey fringes converges towards $\delta_{\rm spectro}^0$ \footnote{We note that the value of $\delta_{\rm spectro}^0$ on Fig.~\ref{fig:ramsey} is lower than the shift in the linear regime predicted on Fig.~\ref{fig:freq_vs_spacing} because the experimental parameters are slightly different due to the absence of the lattice for these measurements.}. This allows to understand the observations of Fig.~\ref{fig:ramsey}.\\ 

This data establishes a connection between the two regimes of the collective Lamb shift: in Ramsey spectroscopy the shift depends on the amount of excitation in the system, but as time evolves the shift converges to the low-excitation limit due to the decay of the excited state, and one recovers the shift measured in linear-optics fluorescence spectroscopy. The time     dependence of the collective Lamb shift reported here matches the mean-field expectations. For shorter interatomic distances, one expects a departure from the mean-field predictions, opening the possibility to test beyond mean-field theories \cite{Cidrim2021}. \par

\section{Conclusions}
To conclude,
we have demonstrated single-shot atom resolved readout of the atomic state in an atom array collectively interacting with light. As a first demonstration of the possibilities it offers, we measured the excitation fraction with single atom resolution and measured the collective Lamb shift both in the linear and nonlinear optics regime. Our single-shot state readout opens the way for an array of new measurements. In particular, it allows for the measurement of atom-atom correlations as in Rydberg-based experiments \cite{Browaeys2020,Kaufman2021}, so as to measure beyond mean-field effects in situ. For instance, one could investigate atomic correlations induced by collective dissipation \cite{Henriet2019,Masson2020,Facchinetti2016,Plankensteiner2015}
or the steady-state of driven-dissipative spin arrays \cite{Olmos2014,Parmee2017,Parmee2020,Zhang2024}. The addition of local addressing could allow to directly prepare super- or subradiant states \cite{Ballantine2021,Rubies2022,Fayard2023}.


\begin{acknowledgments}
This project has received funding by the Agence Nationale de la Recherche (JCJC grant DEAR, ANR-22-PETQ-0004 France 2030, project QuBitAF), by the European Union (ERC StG CORSAIR,  101039361, ERC AdG ATARAXIA 101018511), and the Horizon
Europe programme HORIZON-CL4- 2022-QUANTUM-02-
SGA (project 101113690 PASQuanS2.1).\\
All of the data from this manuscript is available on \cite{zenodo}.
\end{acknowledgments}
\setcounter{figure}{0}
\renewcommand\thefigure{S\arabic{figure}} 
\setcounter{equation}{0}
\renewcommand\theequation{S\arabic{equation}}

\begin{figure*}[t]
    \includegraphics[scale=0.4]{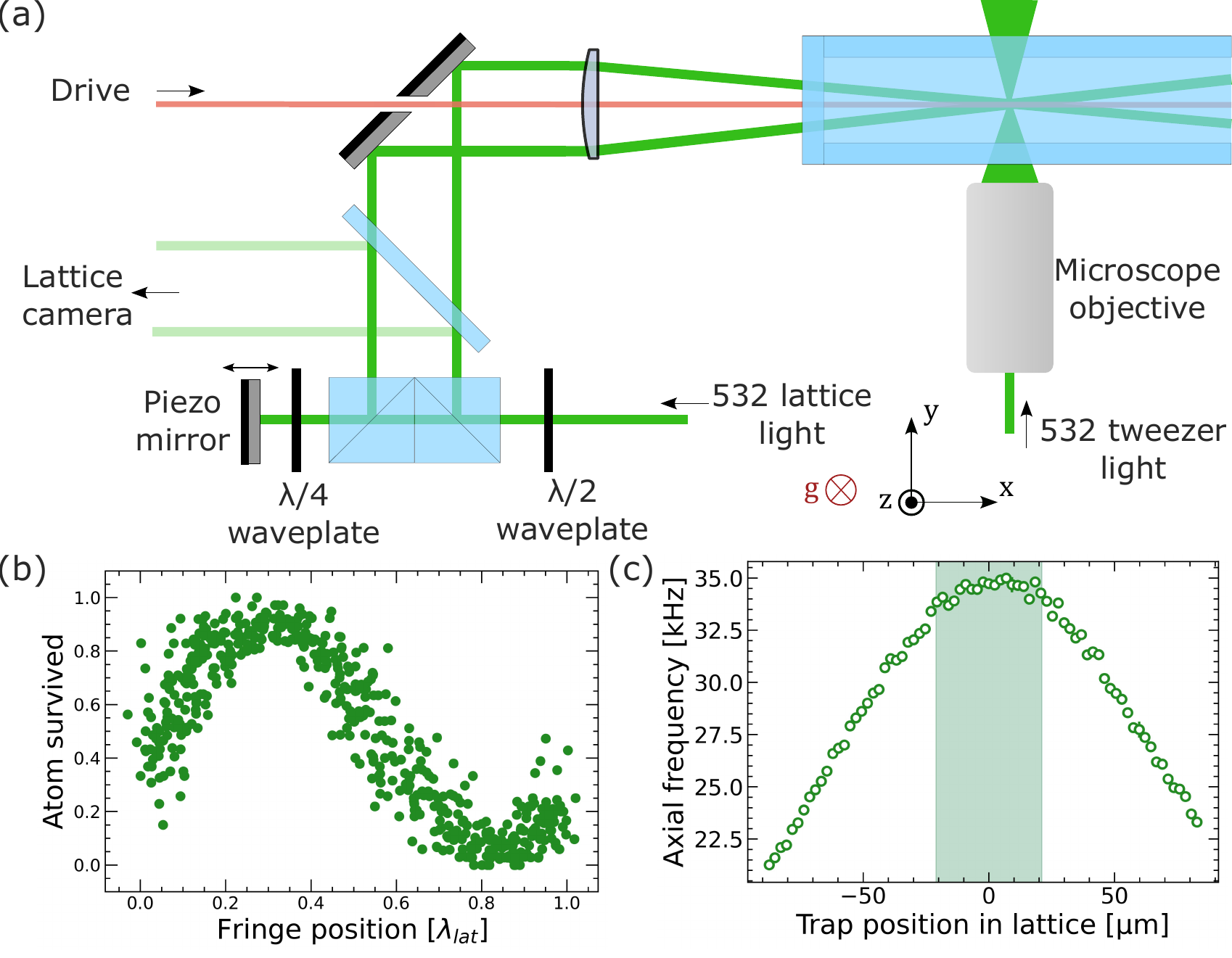}
    \caption{
    \textbf{(a)} View from above of the optical setup for the light sheet creating additional axial confinement in the y direction.
    \textbf{(b)} Atom survival as we abruptly switch the lattice on for different positions of the bright fringe on the atoms. 
    \textbf{(c)} Measurement of the axial trapping frequency across the chain. In order to have a homogeneous axial trapping frequency along the chain we only use the shaded area which corresponds to 30 atoms with a \SI{1.4}{\micro\meter} spacing.
    }
    \label{fig:lattice_setup}
\end{figure*}
\appendix
\section{SHALLOW-ANGLE LATTICE}\label{appendix:axial_trapping}

As presented in the main text, we impose an additional confinement along the tweezers' axial direction by using a shallow angle lattice. This lattice is made by interfering two beams of wavelength $\lambda_l=\SI{532}{\nano\meter}$ at the position of the tweezers (Fig.~\ref{fig:lattice_setup}a). The lattice beams are off resonance from the tweezers' beams by more than \SI{300}{MHz} to avoid any unwanted interference. With this lattice, the axial confinement increases from \SI{7}{kHz} up to \SI{35}{kHz} on the central lattice sites where we perform the experiment. The lattice beams' waist is \SI{10}{\micro\meter} at the atoms' position. The half-angle between the two beams is $\alpha = 5^\circ$ yielding a lattice spacing of $\lambda_l/2\sin\alpha=\SI{3}{\micro m}$. The lattice angle and beam waist (Rayleigh length $\approx600\,\unit{\micro m}$) are chosen to obtain a relatively homogeneous axial trapping frequency across the chain of 30 atoms (see figure\,\ref{fig:lattice_setup}(c)). We load all of the atoms into only one bright fringe of the lattice because the planes are far apart due to the large lattice spacing. To generate the two beams we use two cubes and a mirror on a 3-axis-mount. The retroreflection mirror is mounted on a piezo stack to control the phase difference between the beams and place the central fringe on the atoms. The position of the lattice fringe is stabilized by monitoring the fringes on a camera. The lattice beams are sampled shortly before the atoms (see Fig.~\ref{fig:lattice_setup}(a)). The relative phase between the beams on the camera and the beams on the atoms nevertheless drifts, resulting in a slow drift of the position of the bright fringe. To monitor this, we measure the fringe position directly on the atoms in the following way. At the end of every experimental sequence we abruptly turn the lattice on and off, the atoms only survive this lattice pulse when the bright fringe is on the atom since it exerts no center-of-mass kick (see Fig.~\ref{fig:lattice_setup}(b)). By tracing the atoms' survival to this lattice pulse over time we can determine if the position of the bright fringe slowly drifts and we then correct for it. \par 



\section{SINGLE-SHOT STATE READOUT}\label{appendix:readout}

\begin{figure}
    \includegraphics[width=\columnwidth]{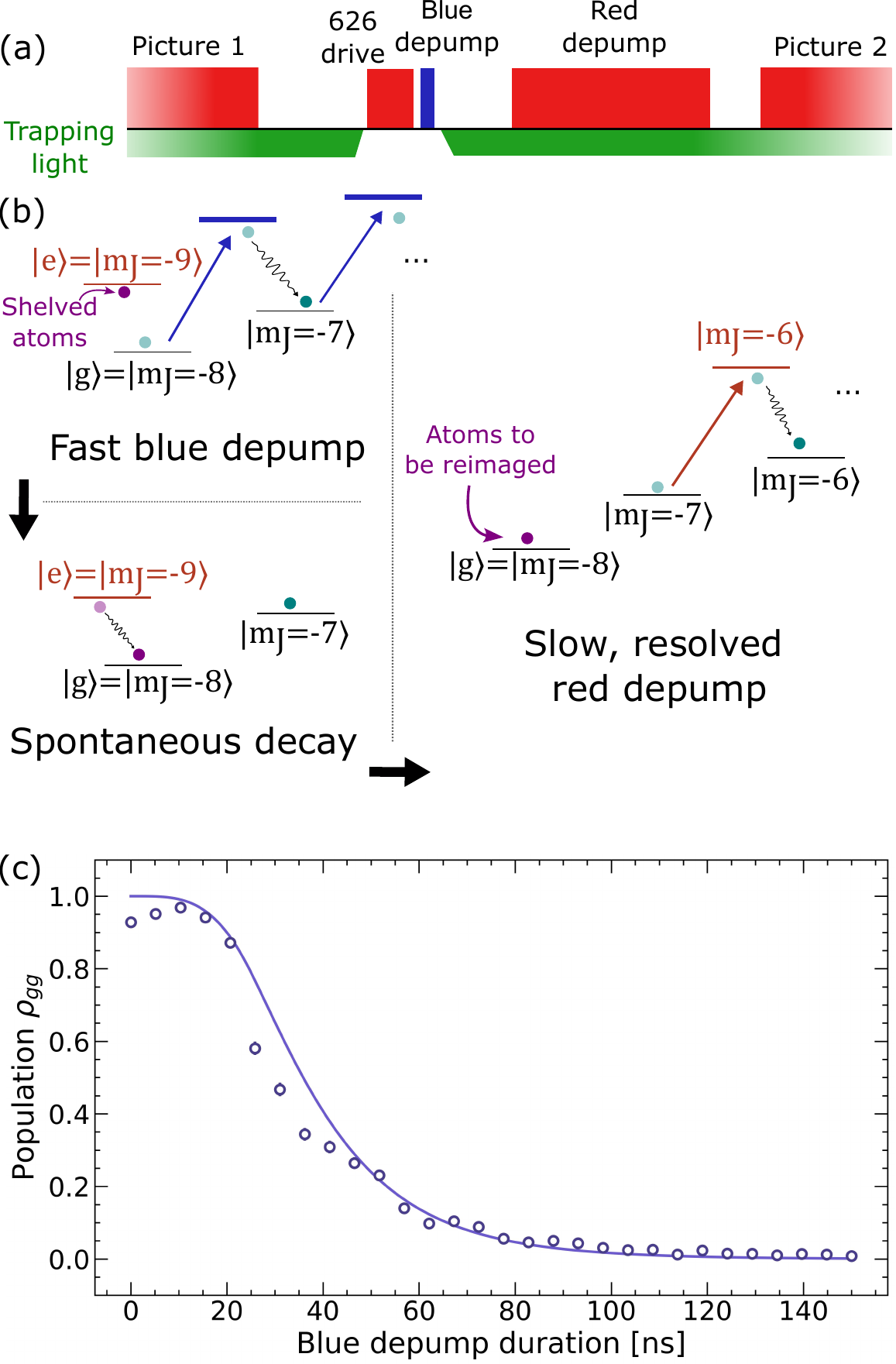 }
    \caption{\textbf{(a)} Pulse sequence for state readout. The experiment is done in time of flight in order to eliminate any systematic errors we may have due to the traps.  \textbf{(b)} Internal state dynamics during state readout: atoms in $\ket g$ are depumped by a fast pulse of $\pi$-polarized light on the broad \SI{421}{nm} transition. Following this pulse, atoms that were in $\ket g$, end up in higher Zeeman states ($m_J>-8$) while atoms that were in $\ket e$ have decayed to $\ket g$. We apply a second pulse of \SI{626}{nm} $\sigma^+$-polarized light that selectively further depumps atoms in $m_J=-7$ to prevent unintentional further repumping during imaging of the $\ket g$ atoms. \textbf{(c)} Calibration of the depumping time. We apply a \SI{421}{nm} pulse of varying duration, atoms are all in $\ket g$ initially, we show the ratio of atoms which survive this pulse. We also plot the solutions to the optical Bloch equations for the 
    $J=8\to J=9$ transition (36 states) for the evolution of the population of an atom starting in the $\ket{g}=m_J=-8$ state and subject to a pulse of \SI{421}{nm} light, taking into account the rise time of the EOM we use to turn the blue light on.
    }
    \label{fig:pulses}
\end{figure}

 In a typical experimental sequence, one atom scatters on the order of one to a few photons during the drive. Given a typical collection efficiency (a few \%), it is impossible to measure the scattered photons in a single shot. Instead, we directly measure the internal state of each atom at a given time. This can be done because the transition that we use has a linewidth of $\Gamma_{626}=2\pi \times \SI{135}{kHz}$ \emph{i.\,e.\,}a lifetime of $\ket e$ of $\SI{1.2}{\micro\second}$ which is relatively long. In particular it is much longer than the time it takes to scatter a few photons on the broad transition ($\Gamma_{421}=2\pi \times \SI{32}{MHz}$). We apply a \SI{100}{ns} pulse of \SI{421}{nm} light. To generate this pulse, we use an AOM with a \SI{150}{ns} rise time and an extinction greater than 10000 followed by a Pockels cell EOM with an extinction greater than 5000 and with a \SI{10}{ns} rise time. The light has linear polarization along the magnetic field axis to excite the $\pi$ transition. The atoms in $\ket g$ are depumped by the light pulse to other Zeeman states $\ket{d}=\ket{J=8;m_J>-8}$ (see figure\,\ref{fig:pulses} (b)), these atoms are then not re-imaged in subsequent images. The $\ket e$ atoms are shelved from the blue pulse, and then decay down to $\ket g$, so that they are re-imaged in the second picture. We have observed that atoms that are depumped to $m_J=-7$ but not further might actually be repumped by the imaging light and appear as $\ket e$ atoms. To prevent this, we add an extra depumping stage once all the atoms are in the ground state manifold (the sequence of pulses can be seen on Fig.~\ref{fig:pulses} (a)). It is performed with $\sigma^+$-polarized light on the $\SI{626}{nm}$ transition. This transition is sufficiently narrow to be selective and depumps atoms in $m_J=-7$ to higher Zeeman states without impacting atoms in $m_J=-8$ (that were originally in $\ket e$).\par
 The fidelity of this state readout is limited by the lifetime of $\ket e$. During the \SI{100}{ns} pulse, a fraction of $\ket e$ atoms decay to $\ket g$, limiting the fidelity of the state readout. In addition, some $\ket g$ atoms might remain in $\ket g$. 
 This leads to a probability of measuring an atom in a given state: 
\begin{equation}
    P(m_x)=P(m_x|x)P(x)+P(m_x|y)P(y)
    \label{eq:prob}
\end{equation}
Where $P(m_x)$ is the probability to measure the atom in state $\ket x$ ($x,y=e,g$), $P(m_x|y)=1-P(m_y|y)$ is the probability of measuring the atom in $\ket{x}$ while it was actually in $\ket{y}$ and $P(x)$ is the probability for the atom to be in $\ket{x}$. On the experiment, we calibrated $P(m_e|g)$ and $P(m_g|e)$ in the following manner:
First, we start with all atoms in $\ket{g}$, we then apply the blue pulse. The fraction of atoms that were not depumped and that we re-image after this gives us $P(m_e)=P(m_e|g)=0.05$. We show in Fig.~\ref{fig:pulses}(c) how $P(m_e)$ evolves with the time of the blue pulse. In blue solid line, we show the theoretical expectation. This is obtained by calculating the population in $\ket g= \ket{J=8,m_J=-8}$ by solving the master equation of the $J=8\to J'=9$ system driven by blue light with $\pi$ polarization and taking into account the rise time of the Pockels cell EOM. The best fit (solid blue line) is obtained by taking a Rabi frequency of $\Omega_{421}=\SI{50}{MHz}$ for the \SI{421}{nm} light. 

Second, we prepare an incoherent mixture of $\ket g$ $\ket e$ ($P(e)=P(g)=0.5$) by sending a resonant pulse of \SI{626}{nm} light with a high Rabi frequency $\Omega=2\pi\times\SI{3}{MHz}$ for a time $t\gg1/\Gamma$. We measure $P(m_e)$ and then determine the fidelity, we get $P(m_e|e)=0.92$. This is in good agreement with our expectations since the fraction of population initially in $\ket e$ that decays during the blue pulse and might be depumped is expected to be $e^{-\tau_p/\tau}=0.92$. We show observed Rabi oscillations in the main text (Fig.~\ref{fig:figure1}(c)) when driving the $\ket g\to\ket e$ transition with the raw data in blue dots and the data corrected for detection errors using the above calibration in red diamonds. The data are very well fitted by a solution of the optical Bloch equations \cite{steck2007}, with Rabi frequency $\Omega=\SI{3.6}{MHz}$ and a transverse decay rate $\gamma_\perp = \SI{60}{kHz}$ (with $\Gamma = (2\pi)\,\SI{135}{kHz}$). We assign this transverse decay rate to laser phase noise and inhomogeneity of the Rabi frequency among the different atoms. 


\section{SINGLE-ATOM EXCITATION PROBABILITY ANALYSIS}\label{appendix:data analysis}
The data plotted in Fig.~\ref{fig:excitation_prob} is an average of many experimental realizations, sometimes separated by a few weeks. We therefore have to eliminate systematic effects. First, residual inhomogeneities of the site resolved excitation imaging along the chain. For this we calibrate $P^{(i)}(m_e|e)$ and $P^{(i)}(m_g|g)$ at each site $(i)$. We also measure the effect of switching the tweezers off during the red driving light and the blue depump pulse. We measure $P_S^{(i)}$ the probability for atom $i$ to survive the switch on and off of the tweezers in absence of the drive and depump beams. These quantities are calibrated by interleaved measurements during the data taking. What is finally plotted in Fig.~\ref{fig:excitation_prob} is $P^{(i)}(e)/P_{S}^{(i)}$. This allows us to eliminate any systematic effects which could be non homogeneous along the atomic chain.  Furthermore, the driving laser's Rabi frequency was not constant over all datasets. We accounted for it by normalizing the data by the measured probability of the first atom to be excited. We note that since this individual point is prone to statistical noise, this results in a statistical offset of all points, but we use this to keep the data analysis free of adjustable parameters. \\

\section{MEAN-FIELD SIMULATIONS}\label{appendix:NLCD}
We perform simulations of the light scattering experiments based on the mean-field equations that can be derived from the full master equation. They are derived in many works, see for instance \cite{doEspirito2020,Glicenstein2020} for more details. 

They read:
\begin{equation}
    \begin{cases}
      \dv{\rho_{ee,n}}{t}=-\Gamma\rho_{ee,n}+\frac{i}{2}\left(\Omega_n\rho_{eg,n}^*-\Omega_n^*\rho_{eg,n}\right)\\
      \dv{\rho_{eg,n}}{t}=-\left(\frac{\Gamma}{2}-i\Delta_L\right)\rho_{eg,n}+\frac{i}{2}\Omega_n\left(1-2\rho_{ee,n}\right)
    \end{cases}
    \label{eq:OBE}
\end{equation}
$\Omega_n$ is the total Rabi frequency for atom $n$. It has two contributions: the drive, and that of the other dipoles in the chain (called $\Omega_\mathrm{atoms}$ in the main text). It is written in terms of the fields:
\begin{equation}
    \Omega_n=\frac{d_0}{\hbar}\vec{\epsilon}^*\cdot \vec{E}_{tot}=\frac{d_0}{\hbar}\vec{\epsilon}^*\cdot \left(\vec{E}_L\left(\vec r_n\right)+\sum_{k\neq n}\vec{E}_k\left(\vec{r}_n-\vec r_k\right)\right)
    \label{eq:omega_tot}
\end{equation}
with $\vec\epsilon=(\hat y-i\hat z)/\sqrt2$ the polarization of the transition we consider here: $\sigma^-$-transition with magnetic field along $\hat x$. $\vec{E}_L$ is the laser field and $\vec{E}_k\left(\vec{r}_n-\vec r_k\right)$ is the field radiated by atom $k$ on atom $n$. 
The field of atom $k$ at the position of atom $n$ is $\vec{E}_k\left(\vec{r}_n- \vec r_k\right)= d_k\overleftrightarrow{\textbf{G}}(\vec{r}_n-\vec{r}_k)\cdot\vec{\epsilon}$ where $d_k=2\rho_{eg}d_0$ is the dipole of atom $k$, $d_0=\bra{e}\vec \epsilon^*.\vec{\hat{d}}\ket{g}=\sqrt{\frac{3\pi\epsilon_0\Gamma\hbar}{k^3}}$ is the dipole matrix element and the Green's function:
\begin{align*}
    \overleftrightarrow{\mathbf{G}}(\mathbf{r}) &= \frac{k^3}{4\pi\epsilon_0}e^{ikr}\left(\frac{1}{kr}+\frac{i}{(kr)^2}-\frac{1}{(kr)^3}\right)\mathbb{I}\\
    &+\frac{k^3}{4\pi\epsilon_0}e^{ikr}\left(-\frac{1}{kr}-\frac{3i}{(kr)^2}+\frac{3}{(kr)^3}\right)\ket{\hat{\mathbf{r}}}\bra{\hat{\mathbf{r}}}.
     \label{eq:greens_func}
\end{align*}
The 30 atoms of the chain have positions $\vec{r}_i=\vec{r}_{0,i}+\vec{\delta r}_i$, where $(\vec{r}_{0,i})_{i\in [0, N-1]}$ are the positions of the 30 tweezers perfectly spaced by $d$. To average on positional disorder due to temperature, we draw the random positions $\vec{\delta r}_i$ following the Boltzmann distribution $\exp\left({-\frac{m\omega^2r^2}{2k_BT}}\right)$. We solve the set of equations \ref{eq:OBE} taking into account the rise time of the AOM and the time scale during which we do the experiment and obtain simulated spectra which we use to extract the shift. 
We have further verified in the linear regime that accounting for Doppler broadening due to thermal motion does not lead to a significant reduction of the calculated shifts. \\
\section{ANALYTICAL FORMULA FOR THE FREQUENCY SHIFT}\label{appendix:analytical_shift}
Here we analytically derive the lineshape in the weakly interacting case. We show that it is a skewed Lorentzian with a shift that we define. Let us start with the master equation for the density matrix $\frac{\mathrm{d}\hat{\rho}}{\mathrm{d}t}=\frac{1}{i\hbar}\left[\hat{H},\hat{\rho}\right]+\mathcal{L}[\hat{\rho}]$ where the Hamiltonian is \cite{Asenjo2017}:
\begin{equation*}
      \hat{H}/\hbar=\sum_i\left[-\Delta_L\hat{\pi}_i^e+\frac{\Omega_i}{2}\hat{\sigma}_i^+\frac{\Omega_i^*}{2}\hat{\sigma}_i^-\right]+\sum_{i\neq j}J_{ij}\hat{\sigma}_i^+\hat{\sigma}_j^-
    \label{eq:hamiltonian}
\end{equation*}
and the dissipation is represented by the operator $\mathcal{L}$:
\begin{align*}
          \mathcal{L}[\hat{\rho}]&=\Gamma\sum_i\left[\hat{\sigma}_i^-\hat{\rho}\hat{\sigma}_i^+-\frac12\{\hat{\sigma}_i^+\hat{\sigma}_i^-,\hat{\rho}\}\right]\\
          &+\sum_{i\neq j}\Gamma_{ij}\left[\hat{\sigma}_i^-\hat{\rho}\hat{\sigma}_j^+-\frac{1}{2}\{\hat{\sigma}_i^+\hat{\sigma}_j^-,\hat{\rho}\}\right]
\end{align*}
The coupling terms are defined as $J_{ij}=\Re[V_{ij}]/\hbar$ and $\Gamma_{ij}=-2\Im[V_{ij}]/\hbar$. Where we have simplified the notation for the dipole dipole interaction as $V_{ij}=V_{dd}(\vec r_i-\vec r_j)=-d_0^2\ \vec \epsilon^*.\overleftrightarrow{\textbf{G}}(\vec r_i-\vec r_j).\vec \epsilon$. The expectation value of an operator $\hat{O}$ is $\langle\hat{O}\rangle={\rm Tr}\{\hat{O}\hat{\rho}\}$ and we write the time derivative of the expectation value of the population operator for the excited state $\hat{\pi}_i=|e_i\rangle\langle e_i|$ of atom i, making a mean-field approximation:
\begin{align*}
    \frac{\mathrm{d}\langle\hat{\pi}_i^e\rangle}{\mathrm{d}t}&=i\frac{\Omega_i}{2}\langle\hat{\sigma}_i^-\rangle-i\frac{\Omega_i^*}{2}\langle\hat{\sigma}_i^+\rangle-\Gamma\langle\hat{\pi}_i^e\rangle\\
    &+\frac{i}{\hbar}\sum_{j\neq i}\left[V_{ij}^*\langle\hat{\sigma}_j^+\rangle\langle\hat{\sigma}_i^-\rangle-V_{ij}\langle\hat{\sigma}_j^-\rangle\langle\hat{\sigma}_i^+\rangle\right]
    \label{eq:dpi_dt}
\end{align*}
We do the same for the lowering operator for atom i $\hat{\sigma}_i^-=|g_i\rangle\langle e_i|$:
\begin{equation}
    \begin{split}
        \frac{\mathrm{d}\langle\hat{\sigma}_i^-\rangle}{\mathrm{d}t}&=(i\Delta_L-\Gamma/2)\langle\hat{\sigma}_i^-\rangle+i\frac{\Omega_i}{2}(2\langle\hat{\pi}_i^e\rangle-1)\\
        &+\frac{i}{\hbar}\sum_{j\neq i}V_{ij}\langle\hat{\sigma}_j^-\rangle(2\langle\hat{\pi}_i^e\rangle-1)
        \label{eq:d_sigma_dt}
    \end{split}
\end{equation}

We then solve this set of coupled equations in steady state to first order in $V_{ij}$ and we obtain the average excited state population in the array $\rho_{\rm ee}$ summing over all atoms:
\begin{widetext}
\begin{eqnarray}
     \rho_{\rm ee}=\frac1N\sum_i\langle\hat{\pi}^e_i\rangle&\approx& \frac{\Omega^2}{1+s(\Delta_L)}\left[1+4\frac{(1+4\Delta_L^2/\Gamma^2)(1+2\Omega^2/\Gamma^2)}{(1+s(\Delta_L))^2}\left(2\frac{\Delta_L}{\Gamma}\delta_\mathrm{spectro}+\frac{\gamma_\mathrm{spectro}}{\Gamma}\right)\right]\label{eq:line}\\
        s(\Delta_L)&=&4\Delta_L^2/\Gamma^2+2\Omega^2/\Gamma^2,
\end{eqnarray}
with $\delta_\mathrm{spectro}=\delta_\mathrm{spectro}^0/(1+2\Omega^2/\Gamma^2)$, $\gamma_\mathrm{spectro}=\gamma_\mathrm{spectro}^0/(1+2\Omega^2/\Gamma^2)$, and 
\begin{eqnarray*}
    \delta_\mathrm{spectro}^0=\frac{1}{N\,\hbar}\sum_{n\neq m}\mathrm{Re}[V_\mathrm{dd}(\vec r_n-\vec r_m)e^{-i\vec k.(\vec r_n-\vec r_m)}]\\
    \gamma_\mathrm{spectro}^0=\frac{1}{N\,\hbar}\sum_{n\neq m}\mathrm{ Im}[V_\mathrm{dd}(\vec r_n-\vec r_m)e^{-i\vec k.(\vec r_n-\vec r_m)}]
\end{eqnarray*}
\end{widetext}
In the limit where the shift is small, we find that the derivative of $\langle\hat{\pi}_i^e\rangle$ cancels out for a laser detuning equal to $\delta_\mathrm{spectro}$. When analyzing the experimental data and results of mean-field simulations, we fit the results with Eq.\,\eqref{eq:line}, leaving $\delta_\mathrm{spectro}$ as a free parameter. 
\\
\section{TIME-DEPENDENT SHIFT IN RAMSEY SPECTROSCOPY}\label{appendix:ramsey_shift}
To derive the evolution of the frequency shift when doing Ramsey spectroscopy, we analytically calculate the evolution of the expectation values of the Bloch vectors $\hat{\sigma}_i^X=\hat \sigma_i^++\hat \sigma_i^-$ and $\hat{\sigma}_i^Y=i(\hat \sigma_i^--\hat \sigma_i^+)$ and $\hat{\sigma}_i^Z=2\hat \pi_i^e-1$. In order to remove the position dependence on the atoms we introduce $\tilde \sigma_i^{-(+)}=e^{-i\vec k_\mathrm{las}.\vec r_i}\hat \sigma_i^{-(+)}$, with $\vec k_\mathrm{las}$ the excitation laser wavevector. For this derivation we use perturbation theory and approximate the solution of the expectation values of the Bloch vectors by $\langle\tilde \sigma_i^{-(+)}\rangle\approx \langle\tilde \sigma_i^{-(+)}\rangle_{NI} + \langle\delta\tilde \sigma_i^{-(+)}\rangle  V_{ij}$ where $\langle\tilde \sigma_i^{-(+)}\rangle_{NI}$ is the Bloch vector in the non-perturbed case. We start by expressing the Bloch vectors in the non-interacting case by using the equations describing the Bloch vector evolution \cite{steck2007}:
\begin{equation*}
    \begin{cases}
        \frac{\mathrm d\langle\tilde \sigma_i^X\rangle}{\mathrm dt}=\Delta_L\langle\tilde \sigma_i^Y\rangle-\frac{\Gamma}{2}\langle\tilde \sigma_i^X\rangle\\
        \frac{\mathrm d\langle\tilde \sigma_i^Y\rangle}{\mathrm dt}=-\Delta_L\langle\tilde \sigma_i^X\rangle-\frac{\Gamma}{2}\langle\tilde \sigma_i^Y\rangle-\Omega\langle\tilde \sigma_i^Z\rangle\\
        \frac{\mathrm d\langle\tilde \sigma_i^Z\rangle}{\mathrm dt}=-\Gamma\left[\langle\tilde \sigma_i^Z\rangle+1\right]+\Omega\langle\tilde \sigma_i^Y\rangle\\
    \end{cases}
    \label{eq:Bloch_vectors}
\end{equation*}
First, a pulse of area $\theta_0$ is applied which rotates the Bloch vector around the x-axis. The system then evolves freely for a time $T_\mathrm{R}$. We solve equations \ref{eq:Bloch_vectors} for $\Omega=0$ and find that:
\begin{equation*}
    \begin{cases}
        \langle\tilde \sigma_i^X\rangle_{NI}=\mathrm{sin}(\theta_0)\mathrm{sin}(\Delta_LT)e^{-\Gamma T_\mathrm{R}/2}\\
        \langle\tilde \sigma_i^Y\rangle_{NI}=\mathrm{sin}(\theta_0)\mathrm{cos}(\Delta_LT)e^{-\Gamma T_\mathrm{R}/2}\\
        \langle\tilde \sigma_i^Z\rangle_{NI}=(1-\mathrm{cos}(\theta_0))e^{-\Gamma T_\mathrm{R}}-1
    \end{cases}
\end{equation*}
Using the mean-field Eq.\,\eqref{eq:d_sigma_dt} for $\Omega=0$ and truncating at first order in $\tilde V_{ij}=V_{ij}e^{i\vec k.(\vec r_i-\vec r_j)}$ we get:
\begin{align*}
    \langle\tilde \sigma_i^-\rangle=&\,\mathrm{sin}(\theta_0) e^{(i\Delta_L-\Gamma/2)T_\mathrm{R}}\\
    &\left[1+\frac{i}{\hbar}\sum_{j\neq i}\tilde V_{ij}\int_0^{T_\mathrm{R}}\langle\tilde \sigma_i^Z(\tau)\rangle_{NI} \mathrm d\tau\right]
\end{align*}
Assuming the interactions are small, we can rewrite this equation using the previous definitions for $\delta_\mathrm{spectro}^0$ and $\gamma_\mathrm{spectro}^0$ by using the total dipole $\frac{1}{N}\sum_{i}\langle\tilde\sigma_i^-\rangle=\langle\tilde\sigma^-\rangle=\mathrm{sin}(\theta_0)\,e^{T_\mathrm{R}\left(i\,\alpha+\beta\right)}$, with:
\begin{eqnarray*}
    \alpha = \Delta_L-\delta_\mathrm{spectro}^0 S_\theta(T_\mathrm{R})\\
    \beta = -\frac{\Gamma}{2}+\gamma_\mathrm{spectro}^0 S_\theta(T_\mathrm{R}),
\end{eqnarray*}
This expression shows that the shift of the Ramsey fringes is $\delta_{\mathrm{Ramsey}}(T_\mathrm{R})=\delta_\mathrm{spectro}^0S_\theta(T_\mathrm{R})$, where we defined:
\begin{eqnarray*}
    S_\theta(T)&=&-\frac{1}{T}\int_0^{T}\langle\tilde\sigma^Z(\tau)\rangle_{NI} \mathrm d\tau\\
    &=&1-\left(1-e^{-\Gamma T}\right)\frac{1-\mathrm{cos}(\theta_0)}{\Gamma T}
    \end{eqnarray*}

\bibliography{sample}


\end{document}